\documentclass[12pt]{article}
\usepackage{amsmath,amscd,amssymb,amsthm}
\theoremstyle{plain}
\newtheorem{theorem}{{\bf Theorem}}

\begin{document}

\begin{center}\begin{Huge}{\bf Reliability of swarming algorithms for mobile sensor network applications\\} \end{Huge} Poster Summary for FFT 2012 \begin{Huge}  \vskip .5in Steven Senger\\ \vskip .25in University of Delaware \vskip .25in
\end{Huge}senger@math.udel.edu\end{center}
\section{Abstract}

There are many well-studied swarming algorithms (for example, \cite{CHRE}, \cite{KL}, \cite{LPLSFD}, \cite{MKJR}, \cite{SCLCJS}, \cite{TRS}, and \cite{WWM}) which are often suited to very specific purposes. As mobile sensor networks become increasingly complex, and are comprised of more and more agents, it makes sense to consider swarming algorithms for movement control. We introduce a natural way to measure the reliability of various swarming algorithms so a balance can be struck between algorithmic complexity and sampling accuracy.

The main idea is to utilize relatively well-developed tools (see \cite{BF}) from areas like frame theory to provide theoretical guidance for algorithm selection.
\newpage
\section{Frames, frame potential, and energy}

Frames were first introduced in \cite{DS}. A {\it frame} is a set of vectors which, like a basis of a vector space, span a certain space of interest, but with perhaps (many) more vectors than necessary to span the space. This theory is immediately relevant for transmitting complex information through noisy or lossy media. More precisely, let $\mathbb{H}^d$ be a $d$-dimensional Hilbert space, then the set of vectors $$\mathcal{F} = \lbrace f_j \rbrace_{j=1}^n \subset \mathbb{H}^d,$$
is a frame if there exist constants $A$ and $B$, called {\it frame bounds}, such that for any $y\in \mathbb{H}^d$, the following holds:
$$ A\Vert y \Vert^2 \leq \sum_j |\langle y,f_j \rangle|^2\leq B\Vert y \Vert^2,$$
where $\langle\cdot,\cdot\rangle$ denotes the inner product associated to the Hilbert space.

The more control one has on the frame bounds, the more ``well-behaved" a given frame is. One useful measure of how well a given frame behaves is the so-called {\it frame potential}, introduced in \cite{BF}. The lower the frame potential, the better the frame is suited to many tasks. Retaining the notation above, the frame potential of a given frame is defined to be
$$FP(\mathcal{F})=\sum_{i=1}^n\sum_{j=1}^n|\langle f_i,f_j\rangle|^2.$$

The frame potential is related to a very physically motivated quantity used in many areas of mathematics called {\it energy}. Again, using the same notation as above, the energy of a collection of vectors is defined to be
$$E(\mathcal{F})=\frac{1}{{n \choose 2}}\sum_{i=1}^n\sum_{j>i}^n \Vert \frac{1}{f_i-f_j\Vert^2}.$$
\newpage
\section{Mobile sensor networks}

As the number of agents in mobile sensor networks increases, it will be necessary to consider asymptotic analysis of the algorithms which control them. It stands to reason that tools from frame theory could be usefully employed to study mobile sensor networks by viewing them as frames with dynamic vectors. Here is an example of such a result.

An excellent example would be a swarm of drones flying just above a forest fire, regularly measuring quantities like temperature and humidity. This swarm of sensors is moving coherently in a some direction. Each agent of the swarm is taking samples of some quantity at regular intervals along the way. The agents should be well-distributed, so that the the region under consideration is sampled with minimum bias. Also, whatever quantities are being sampled will be assumed to not change too much, on average, over some small, fixed distance. Notice that the complexity of a given algorithm can strongly influence the possible velocity for such applications.

\begin{theorem}
Let $$\mathcal{X}(t)=\lbrace x_j(t) \rbrace_{j=1}^n \subset \mathbb{R}^d$$ be the positions of agents in a swarm at time $t$. Let $v_j(t)$ denote the velocity of the agent at position $x_j$ at time $t$, and let $$v_{avg}=\frac{1}{n}\sum_{j=1}^n v_j.$$ Suppose further that the desired resolution in space of the sampling is such that each agent has a minimum separation of $\delta>0$ from each other agent, and that $|v_j(t)-v_{avg}|<c\delta$, for some small constant $c$, independent of $n$. Finally, let $\rho(t)$ denote the ratio of the area which is covered by $\delta$-balls centered at the agents to the area of the smallest cube containing the swarm at time $t$. Then for all times $t$ such that
$$E(\mathcal{X}(t))=\frac{1}{{n \choose 2}}\sum_{i=1}^n\sum_{j>i}^n \frac{1}{\Vert x_i(t)-x_j(t) \Vert^2} \leq C,$$
for some value $C$, independent of $n$, $\rho(t) \geq C'$, for some other value $C'$, which is also independent of $n$.
\end{theorem}

\newpage

\section{Conclusion}

Energy can be determined by simulation before any swarm is given a navigation algorithm, so that various parameter sets can be tested cheaply. If a given algorithm or parameter set has energy which depends on the number of agents, there could be an uneven distribution of sensors. However, the theorem above guarantees that the distribution of sensors will be asymptotically robust.

\end{document}